\documentstyle{npbproc}

\def\GeV{G\eV}
\def\sigpiN{\sigma_{\pi N}}
\def\sigpiNv{\sigma_{\pi N}^{\rm val}}
\newbox\Strutbox
\setbox\Strutbox=\hbox{\vrule height8pt depth4pt width0pt}
\def\Strut{\relax\ifmmode\copy\Strutbox\else\unhcopy\Strutbox\fi}

\begin{document}
\title{SCALAR AND AXIAL MATRIX ELEMENTS OF THE NUCLEON: SEA QUARK CONTENT}

\author{Apoorva Patel}
\address{CTS and SERC, Indian Institute of Science, Bangalore-560012, India}

\date{}

\runtitle{Scalar and Axial Matrix Elements of the Nucleon: Sea Quark Content}
\runauthor{A. Patel}

\volume{XXX}  
\firstpage{1} 
\lastpage{3}  

\begin{abstract}
Sea quark contributions to the scalar density and the axial current
matrix elements of the nucleon are studied in lattice QCD with two
flavours of dynamical Wilson fermions. The results are compared to
trends in heavy quark mass expansions, and contrasted with the numbers
obtained using dynamical staggered fermions.
\end{abstract}

\maketitle

In the numerical study of many hadronic properties, a comparison of
the full QCD and the quenched approximation results shows that the
differences are small (though the bare gauge couplings in the two cases
are quite different); small enough that given the various sources of
errors it is hard to differentiate between the two sets of numbers.
Also the phenomenological quark models, which ignore the sea quarks
altogether, do a good job of fitting many of the experimental results.
All this suggests a welcome phenomenological simplification of the
theory (i.e. the dominant effect of the sea quarks is to generate
constituent quarks as valence quarks with renormalised properties),
but one would like to understand the dynamical reason behind it.

To find unambiguous signatures of the sea quarks (or failures of the
quenched approximation), one has to look for instances where the sea quarks
amount to more than mere renormalisation of the gauge coupling. Good places
to search are the properties where the quark model is not a good guide.
Of particular interest are the quark bilinear matrix elements which
naively vanish in the quark model, e.g. the strange quark matrix
elements of the proton $\langle p| \bar s \Gamma s |p \rangle$.
We concentrate on the scalar ($\Gamma=1$) and the axial vector
($\Gamma=i\gamma_\mu\gamma_5$) bilinears. Both are related to
dynamically broken classical symmetries of QCD, and hence
have scope for a significant mixing with gluonic sector/sea quarks.
Moreover, both correspond to Hermitian operators which are numerically
easy to deal with, e.g. using the Hybrid Monte Carlo algorithm.
(Complex fermion determinants are difficult to handle as exemplified
by the attempts to study QCD at finite chemical potential
($\Gamma = i \gamma_0$) and the $\theta-$dependence of the
neutron electric dipole moment ($\Gamma = i \gamma_5$).)

The non-singlet quark bilinear matrix elements of the nucleon,
conventionally expressed in terms of $F$ and $D$ couplings, are easily
extracted from a direct evaluation of the $3-$point correlators.
On the contrary, correlators containing a purely gluonic intermediate
state are quite noisy. Experience has shown that instead of a direct
evaluation of the $3-$point correlators, it is easier to compute the
sea quark matrix elements by making hadrons propagate through a
background external field (i.e. creating sea quarks with an extra
source term $S_\Gamma = \sum_x h_\Gamma \bar{\psi}(x) \Gamma \psi(x)$
added to the standard fermion action) and then evaluating numerical
derivatives with respect to the external field strength $h_\Gamma$.
While a singlet scalar density field is obtained just by choosing
unequal valence and sea quark masses, configurations with a singlet
axial current field have to be created afresh.

\section*{Pion-Nucleon Sigma Term}
The clear and interesting results concern the pion-nucleon sigma term
(with $m=m_u=m_d$) :
\begin{equation}
\sigpiN = \langle N| m (\bar u u + \bar d d) |N \rangle
        = m (\partial M_N / \partial m) ~,
\end{equation}
where the quark mass derivatives are to be evaluated at fixed gauge coupling.
The most recent analysis \cite{sigmaexpt} of experimental data (though
better data are definitely desirable), gives $\sigpiN \approx 45$ \MeV.
Within the first order flavour $SU(3)$ breaking parametrisation,
the valence quark component is only $\sigpiNv \equiv (3F_S-D_S)
\approx 26$ \MeV, leaving ample room for the sea quark component.

We cannot yet directly calculate $\sigpiN$ on the lattice, as it
requires extrapolating the matrix element to very small quark masses.
Instead we get an indication of the importance of insertions on sea
quark loops from the ratio of the full matrix element (valence plus
sea) to its valence part.

We generated $16^4$ lattices with two flavours of dynamical
Wilson fermions and periodic boundary conditions.
These lattices were doubled in the time direction prior to
calculation of hadron spectrum and matrix elements.
We obtained results \cite{qcdwftwo} for $\beta \equiv 6g^{-2}
=5.4,5.5,5.6$, with the quark masses in the range $m_s < m_q < 3m_s$.
Our results for the ratio of the full matrix element
to its valence part are shown in fig.1.
We see that the sea contribution is $1-2$ times the valence part.
This is in qualitative agreement with the experimental data.
The overall magnitude of $m \sigpiNv$ is systematically smaller than
the experimental value, however, an effect possibly due to not having
explored small enough quark masses.

\begin{figure}[t]
  \vspace{7.2cm} \caption{
  The results for sigma term matrix elements with two flavours of
  dynamical Wilson fermions. The curve is the leading term in lattice
  heavy quark mass expansion.}
\end{figure}

A similar analysis can be carried out for the staggered fermion results
of the Columbia group \cite{columbia}. Here the numerical derivatives
for valence quark matrix elements involve large quark mass intervals,
but the consistency between numbers extracted using
$(N,\widetilde\Omega)$ and $(\Lambda,\Xi)$ masses provides a check
on baryon masses being linear in quark masses over the interval.
As table 1 shows, the effect of sea quarks is somewhat smaller in this
case compared to Wilson fermions, but nonetheless significant.
(Note that the effect is a factor of two smaller for heavy fermions.)

The analogous scalar density matrix elements of the $\rho$ and the
$\Delta$ show similar factors between the valence and the full value,
while the magnitude of the matrix elements is roughly proportional to
the number of valence quarks. This suggests a model for constituent
quarks in which the bare quarks are dressed strongly, and in a manner
independent of the hadron that they are in. Indeed, it can be
reasoned \cite{scanom} that the large sea quark component we see
is due to change in the overall scale of the theory; the sea quarks
influence the $\beta-$function through vacuum polarisation.
The really surprising feature of the lattice results then is
(see fig.1) that even relatively heavy sea quarks
(corresponding to the pseudoscalar meson mass up to $1-1.2$ GeV)
give a contribution comparable to the valence part.

\begin{table}[t]
  \caption{
  The sigma term matrix elements with two flavours of dynamical
  staggered fermions at $\beta=5.7$. The baryon states used to calculate
  the matrix elements through numerical mass derivatives are indicated.}
$$\vbox{\tabskip=0pt \offinterlineskip
\halign{\Strut #\ &\Strut #&\vrule #\ &\Strut #\ &\Strut #\ &\Strut #\cr
\ Quark    &$\sigpiN /m$&& \ Quark    &    $\sigpiNv /m$     & $\sigpiNv /m$
\cr
mass range &   ~$(N)$   && mass range
&$(N,\widetilde\Omega)$&$(\Lambda,\Xi)$\cr
\noalign{\hrule}
0.010--0.015 &   ~---    && 0.004--0.010 & 9.0(3.)& 8.5(?)   \cr
0.015--0.020 & 19.2(2.8) && 0.015--0.070 & 8.7(3) & 8.5(1.2) \cr
0.020--0.025 & 11.6(2.3) && 0.020--0.070 & 7.8(2) & 7.8(8)   \cr
             &           && 0.025--0.100 & 7.4(1) & 7.5(3)   \cr
}}$$
\end{table}

\section*{Polarisation of Sea Quarks}
The EMC result on polarised muon-proton scattering is difficult to
digest without a sizeable contribution from the sea quarks.
Lattice measurement of sea quark and gluon components of the structure
function $g_1$ can be attempted in two ways. The simpler approach,
possible even within the quenched approximation, is to replace the
insertion on sea quark loops by an effective gluon operator \cite{ffdlatt}.
We have adopted the alternative approach of directly determining the
axial current coupling to sea quarks in the nucleon \cite{gonelatt}.

In the first trial run, we used two flavours of dynamical Wilson
fermions on $8^4$ lattices at $\beta = 5.3$ with $\kappa = 0.165,0.166$.
It is necessary to have the field strength $h_A$ of the order of the
quark mass to see any signal, and we used $2\kappa h_A = 0.005,0.004$.
The pseudoscalar meson masses are about $0.9,0.8$ \GeV\
for these heavy quarks.
We only find $2\sigma$ bounds for the sea quark contribution to axial
current matrix elements of the nucleon, $|S_A| < 0.05,0.08$ per flavour.
These are a factor of $2$ to $4$ below what is required
to confirm the EMC result on the lattice.
(The lattice one loop and continuum two loop renormalisation
constants for the singlet axial current are finite,
and can be ignored at the present qualitative level.)
This sharply contrasts with the scalar density case, where sea
and valence quark components are comparable at similar quark masses.
It is entirely possible that $S_A$ moves away from zero at
relatively small quark masses only.

Lattice results for non-singlet scalar density and axial current
matrix elements have turned out to be reasonable \cite{nonsinglet}.
Heavy quark limits and constituent quark model expectations are easy to
evaluate for these matrix elements. Putting everything together:
(a) $F_S$, $m S_S$, $F_A$ and $D_A$ are finite in the heavy quark limit
and turn out to be smooth functions of the quark mass, while
(b) $m D_S$ and $S_A$ vanish in the heavy quark limit as $m^{-2}$
and depart significantly from zero only when the quark mass becomes
of the order of $\Lambda_{QCD}$. This suggests that flavour $SU(3)$
breaking effects are probably small in the former case,
though likely to be sizeable in the latter case.

\section*{ACKNOWLEDGEMENTS}
I thank LANL and ITP (Santa Barbara) for hospitality during the initial
stages of this work, Hong Chen for sending me the detailed results
of the Columbia group, and finally the organisers of LAT91 without
whose support I would not have been able to attend this conference.


\end{document}